\documentclass[twoside]{elsart}
\usepackage{amsmath}
\usepackage{amssymb}
\usepackage{epsfig}

\begin{document}

\def\d{\partial}
\def\um{\,\mu{\rm  m}}
\def\mm{\,   {\rm mm}}
\def\cm{\,   {\rm cm}}
\def \m{\,   {\rm  m}}
\def\ps{\,   {\rm ps}}
\def\ns{\,   {\rm ns}}
\def\us{\,\mu{\rm  s}}
\def\ms{\,   {\rm ms}}
\def\nA{\,   {\rm nA}}
\def\uA{\,\mu{\rm  A}}
\def\mA{\,   {\rm mA}}
\def\A {\,   {\rm  A}}
\def\mV{\,   {\rm mV}}
\def\V {\,   {\rm  V}}
\def\fF{\,   {\rm fF}}
\def\pF{\,   {\rm pF}}
\def\GeV{\, {\rm GeV}}
\def\MHz{\, {\rm MHz}}
\def\uW{\,\mu{\rm  W}}
\def\e {\,  {\rm e^-}}

\renewcommand{\labelenumi}{\arabic{enumi}}
\renewcommand{\labelitemi}{-}

\begin{frontmatter}

\title{Pixel Detectors for Particle Physics and Imaging Applications\thanksref{BMFT} }

\thanks[BMFT]{Work supported by the German Ministerium f{\"u}r Bildung,
              Wissenschaft, Forschung und Technologie (BMBF) under contract
              no.~$05 HA8PD1$\ , by the
              Ministerium f{\"u}r Wissenschaft und Forschung des Landes
              Nordrhein--Westfalen under contract no.~$IV\,A5-106\,011\,98$. and
              by the Deutsche Forschungsgemeinschaft DFG}

\author {N.~Wermes\thanksref{NW}}

\thanks[NW]{Physikalisches Institut, Nussallee 12,
               D-53115 Bonn, Germany, Tel.: +49\,228\,73-3533, Fax:
               -3220, email: wermes@physik.uni-bonn.de
           }
\address{Physikalisches Institut der Universit{\"a}t Bonn, Germany}

\hfill\break
\begin{center}
Talk given at the 9th European Symposium on Semiconductor
Detectors, Elmau, Germany, June 2002
\end{center}

\begin{abstract}
Semiconductor pixel detectors offer features for the detection of
radiation which are interesting for particle physics detectors as
well as for imaging e.g. in biomedical applications (radiography,
autoradiography, protein crystallography) or in Xray astronomy. At
the present time hybrid pixel detectors are technologically
mastered to a large extent and large scale particle detectors are
being built. Although the physical requirements are often quite
different, imaging applications are emerging and interesting
prototype results are available. Monolithic detectors, however,
offer interesting features for both fields in future applications.
The state of development of hybrid and monolithic pixel detectors,
excluding CCDs, and their different suitability for particle
detection and imaging, is reviewed.
\end{abstract}


\end{frontmatter}

\newpage
\section*{Introduction}
The requirements on semiconductor pixel detectors for charge
particle detectors in high energy physics compared to those from
imaging can be very different. In particle physics experiments
individual charged particles, usually triggered by other
subdetectors, have to be identified with high demands on spatial
resolution and timing. In imaging applications the image is
obtained by the un-triggered accumulation (integrating or
counting) of the quanta of the impinging radiation, often also
with high demands (e.g. $\gtrsim$1 MHz per pixel in certain
radiography or CT applications). Si pixel detectors for high
energy charge particle detection can assume typical signal charges
collected at an electrode in the order of 5000-10000 electrons
even taking into account charge sharing between cells and detector
deterioration after irradiation to doses as high as $60$ Mrad. In
tritium autoradiography, on the contrary, or in low energy Xray
astronomy the amount of charge to be collected with high
efficiency can be much below $1000$ e. The spatial resolution is
governed by the attainable pixel granularity to a few to about
10$\mu$m at best, obtained with pixel cell dimensions in the order
of $50 \mu$m to $100 \mu$m. The requirements from radiology
(mammography) are similar, while some applications in
autoradiography require sub-$\mu$m resolutions, not attainable
with present day pixel detectors. For applications with lower
demands on the spatial resolution ($\mathcal{O}(10 \mu$m)) but
with demands on real time and time resolved data acquisition,
semiconductor pixel detectors are however attractive.

Thin detector assemblies are mandatory for the vertex detectors at
collider experiments, in particular for the planned linear
$e^+e^-$ collider. While silicon is almost a perfect material for
particle physics detectors, allowing the shaping of electric
fields by tailored impurity doping, the need of high photon
absorption efficiency in radiological applications requires the
study and use of semiconductor materials with high atomic charge,
such as GaAs or CdTe. For such materials the charge collection
properties are much less understood and mechanical issues in
particular those related to hybrid pixels are abound, most notably
regarding the hybridization of detectors when they are not
available in wafer scale sizes. Last but not least the
cost-performance ratio is an important factor to consider if an
imaging application should be commercially interesting.

\section*{Hybrid Pixel Detectors for Particle Physics}
\begin{figure}[htb]
 \centerline{\epsfig{file=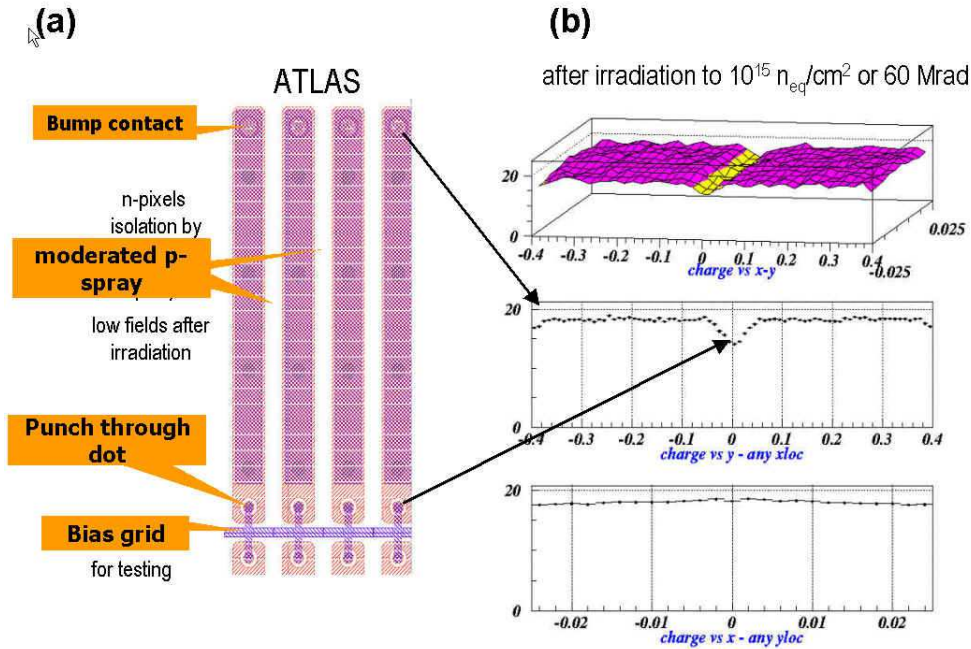,width=130mm}}
  \caption{(a) Layout of the ATLAS pixel sensor, (b) results of
  charge collection measurements obtained in testbeam running with
  detectors and chips which were irradiated to $60$ Mrad.}
  \label{sensors}
\end{figure}

In the "hybrid pixel technique" sensor and FE-chips are separate
parts of the detector module connected by the small conducting
bumps applied by using the bumping and flip-chip technology. All
of the LHC-collider-detectors \cite{ATLAS,CMS,ALICE} ALICE, ATLAS,
and CMS, LHCb for the RICH system \cite{LHCb}, as well as some
fixed target experiments (NA60 \cite{NA60} at CERN and BTeV
\cite{BTEV} at Fermilab) employ the hybrid pixel technique to
build large scale ($\sim m^2$) pixel detectors. Pixel area sizes
are either rectangular (typically $50 \mu$m $\times 400 \mu$m as
for ATLAS) or quadratic ($150 \mu$m $\times 150 \mu$m as for CMS).
The detectors are arranged in cylindrical barrels of $2-3$ layers
and disks covering the forward and backward regions. The main
purpose that these detectors must serve is (a) identification of
short lived particles (e.g. b-tagging for Higgs and SUSY signals),
(b) pattern recognition and event reconstruction and (c) momentum
measurement, in this order of importance. Among the technical
issues of high demand which must be addressed the need to
withstand a total (10 years) particle fluence of $10^{15}$n$_{eq}$
corresponding to a radiation dose of about $60$ Mrad is the most
demanding one. The discovery that oxygenated silicon is radiation
hard with respect to the non-ionizing energy loss of protons and
pions \cite{oxysilicon} saves pixel detectors at the LHC for which
this radiation is most severe due to their proximity to the
interaction point. $n^+$ electrode in n-bulk material sensors have
been chosen to cope with the fact that type inversion occurs after
about $2.5 \times 10^{13}$ n$_{eq}$. After type inversion the
$pn$-diode sits on the electrode side thus allowing the sensor to
be operated partially depleted. Figure \ref{sensors}(a) shows the
layout of the ATLAS pixel sensor \cite{ATLAS-Sensor}. The n$^+$
pixels are isolated against each other by the moderated p-spray
technique \cite{ATLAS-Sensor,RRichter}. The bias grid at the
bottom allowes to test the sensor before bonding to the
electronics ICs is made. All pixels are set under voltage by the
punch through biasing mechanism.
Figure \ref{sensors}(b) show the charge collected in two adjacent
pixels after irradiation with $2 \times 10^{15}$ protons/cm$^2$,
the full LHC dose after 10 years. A tolerable loss in charge
collection efficiency is observed at the punch-through dot, which
goes at the expense of the biasing feature which is too important
to be sacrificed, and at the edge of the pixel. Everywhere else
the collection even after this high dose is homogeneous. The
sensors can be operated after irradiation with bias voltages in
excess of $600$V in full depletion. Pixel sensors for the LHC
detectors are close to or already in production in large
quantities.

The requirements on the FE-chip also impose severe constraints on
noise, threshold dispersion, timewalk and power consumption. After
several chip generations first in radiation soft CMOS and BiCMOS
technologies and later in the radiation hard DMILL and 1/4 micron
technologies have at last produced FE-ICs with the required
performance and decent yield in excess of $60\%$. After
irradiation to the full LHC dose, the ATLAS 1/4 micron pixel chip
FE-I shows noise values of about $250$e and threshold dispersions
in the range of $\sim 70$e after threshold tuning. The power
consumption per pixel is about $50 \mu$W.

\begin{figure}[htb]
 \centerline{\epsfig{file=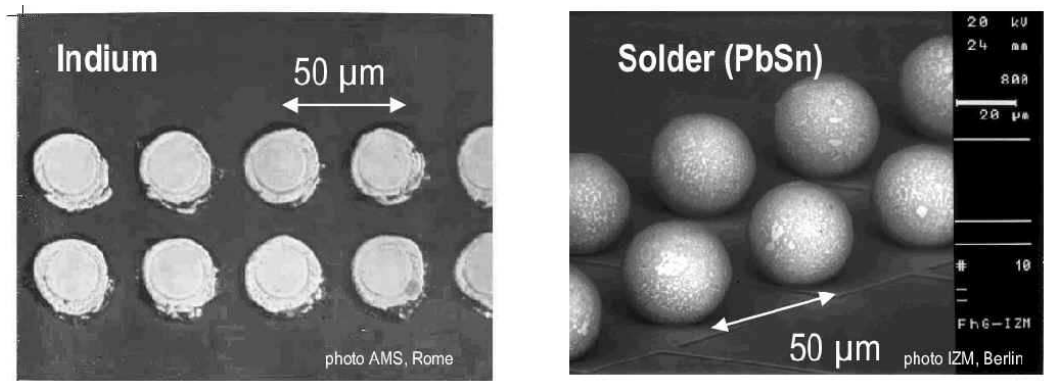,width=140mm}}
  \caption{(left) Indium (Photo AMS, Rome) and (right) solder
  (PbSn, Photo IZM, Berlin) bump rows with $50 \mu$m pitch.}
  \label{bumping}
\end{figure}

The process of chip and sensor connection called hybridization is
done by fine pitch bumping and subsequent flip-chip which is
achieved with either PbSn (solder) or indium bumps at a failure
rate of $< 10^{-4}$. Figure \ref{bumping} shows rows of $50 \mu$m
pitch bumps obtained by these techniques.

Indium bumps are applied using a wet lift-off technique applied on
both sides (sensor and IC) \cite{In-bumping}. The connection is
obtained by thermo-compression. The Indium joint is comparatively
soft and the gap between IC and sensor is about $6 \mu$m. PbSn
bumps are applied by electroplating \cite{PbSn-bumping}. Here the
bump is galvanically grown on the chip wafer only. The bump is
connected by flip-chipping to an under-bump metallization to the
sensor substrate pixel. Both technologies have been successfully
used with 8" IC-wafers and 4" sensor wafers.

\begin{figure}[htb]
 \centerline{\epsfig{file=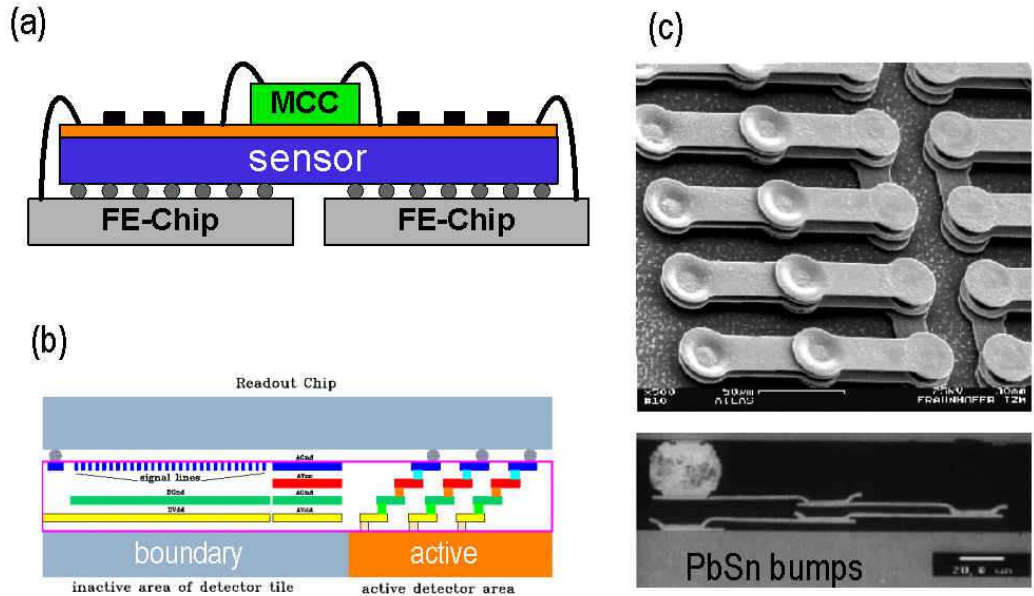,width=140mm}}
  \caption{(a) Schematic view of a hybrid pixel module showing ICs
  bonded via bump connection to the sensor, the flex hybrid kapton
  layer atop the sensor with mounted electrical components,
  and the module control chip (MCC). Wire bond connections are needed as indicated.
  (b) Schematic layout of a MCM-D pixel module. (c) SEM
  photograph of a MCM-D via structure (top) and a cross section
  after bump deposition (bottom).
  }
  \label{module_sketch}
\end{figure}

In the case of ATLAS sixteen FE-chips are bump-connected to a
silicon sensor to form a $module$ of 2.1 cm by 6.4 cm area (fig.
\ref{module_sketch}(a)). The I/O lines of the chips are connected
via wire bonds to a kapton flex circuit glued atop the sensor. The
flex houses a module control chip (MCC) responsible for front end
time/trigger control and event building. The total thickness at
normal incidence is in excess of $2 \%$ of a radiation length.

The modules are arranged in ladders (staves) and cooled by
evaporation of a fluorinert liquid ($C_4F_{10}$ or $C_3F_8$) at an
input temperature below $-20 ^{o}C$ in order to maintain the
entire detector below $-6 ^{o}C$ to minimize the damage induced by
radiation. This operation requires pumping and the cooling tubes
must stand $8$ bar pressure if pipe blocking occurs. All detector
components must withstand temperature cycles between $-25 ^{o}C$
and room temperature.

In summary, the Hybrid Pixel Technology is {\it the} present day
technology for large area pixel detectors in particle physics.
Several pixel vertex detectors are in production for LHC- and
other experiments. Some potentially major issues still exist, most
notably yield issues which in this technology come into play in
many areas as chip-wafer and sensor-wafer yields, bumping and
flip-chipping yields, as well as burn-in yields of modules with 16
ICs and about 50000 individual amplifier channels.

What are the possible future advances and directions and what are
the limitations of the hybrid pixel technology? The area of a
pixel cell is limited by the readout circuitry obtainable for a
given area. With the availability of chip technologies with small
structure sizes ($\leq 0.25 \mu$m) the target pixel size can be
made substantially smaller than planned for the first generation
of large area hybrid pixel detectors. Area sizes of $\sim 50
\times 50 \mu$m$^2$ or somewhat below are, however, at the limit
for this technology in my opinion. The limit for small pitch
bumping is in the order of $10-20$ $\mu$m \cite{Ehrmann}. An
interesting alternative to the flex-kapton solution to provide
power and signal distribution to and from the module is the
so-called Multi-Chip-Module Technology deposited on Si-substrate
(MCM-D), pioneered by IZM (Berlin), for pixel particle detectors
in collaboration with Wuppertal University \cite{MCM-D}. A
multi-conductor-layer structure is built up on the silicon sensor.
This allows to bury all bus structures in four layers in the
inactive area of the module thus avoiding the kapton flex layer
and any wire bonding at the expense of a small thickness increase
of $0.1 \% \ X_0$. Figure \ref{module_sketch}(b) illustrates the
principle and fig. \ref{module_sketch}(c) shows scanning electron
microphotographs (ref. IZM, Berlin) of a via structure made in
MCM-D technology.

\section*{Hybrid Pixel Detectors for Imaging Applications}
\subsection*{Radiology}

There is a vast amount of radiology detection and imaging
techniques. The discussion in this paper will be limited to an
application which possibly opens new directions in radiology due
to fully digital imaging, i.e. pixel detectors with individual
X-ray photon counting in every pixel cell. This approach offers
many features which are very attractive for X-ray imaging:
excellent linearity and an infinite dynamic range (at least in
principle), optimal exposure times and a good image contrast
compared to conventional film-foil based radiography thus avoiding
over- and underexposed images. There are very interesting
developments using integrating methods such as flat panel imagers
based on a-Sci \cite{flat-panel} or a-Se technologies \cite{a-Se},
which are to be considered the state of the art in large area
radiography. Their review, however, is beyond the scope of this
paper and is referred to dedicated conferences \cite{SPIE}.

Two counting hybrid pixel detector developments are called MEDIPIX
\cite{MEDIPIX1}, a CERN-based collaboration and MPEC \cite{MPEC1}
at Bonn University. The challenges to be addressed in order to
become competitive with integrating systems are: high speed
($\gtrsim 1$ MHz) counting with at least 15 bit, operation with
very little dead time, low noise and particularly low threshold
operation with small threshold dispersion values. In particular
the last item is important in order to allow homogeneous imaging
of soft Xrays. It is also mandatory for a differential energy
measurement, realized so far as a double threshold with energy
windowing logic \cite{MPEC-windowing,MEDIPIX2}. A differential
measurement of the energy, exploiting the different shapes of Xray
spectra behind for example tissue or bone, can enhance the
contrast performance of an image. The idea of using a linear
feedback shift register as a small counter which fits in a pixel
area was first realized in \cite{counter} and is implemented in
both, MPEC and MEDIPIX, circuits. Finally, for radiography high
photon absorption efficiency is mandatory, which renders the not
easy task of high Z sensors and their hybridization necessary.

\begin{figure}[htb]
 \centerline{\epsfig{file=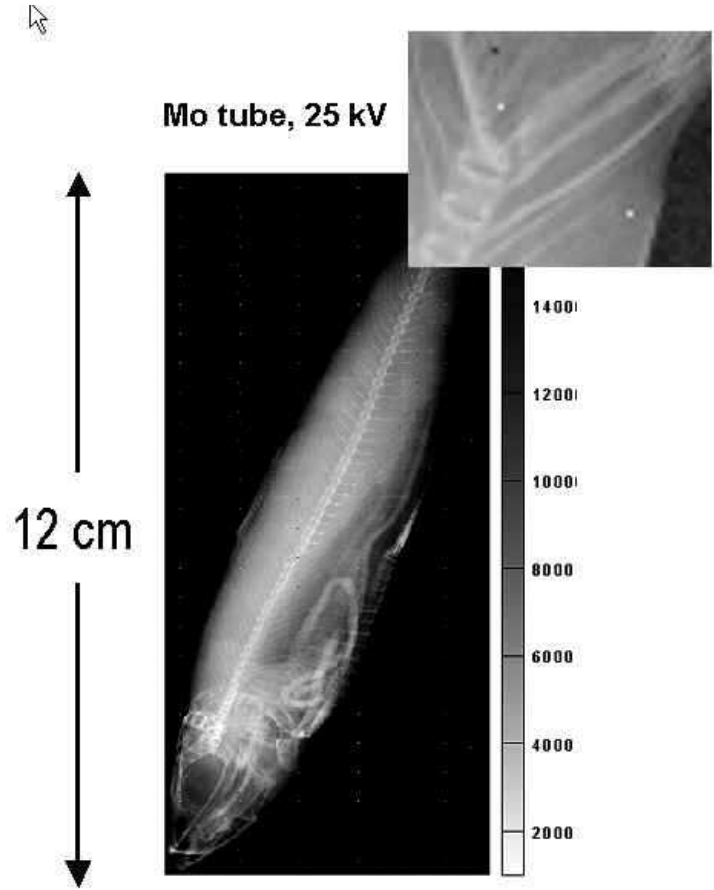,width=100mm}}
  \caption{Image of a sardine obtained with the MEDIPIX1 counting pixel chip
  with Si sensor obtained by successive scans.}
  \label{sardine}
\end{figure}

The MEDIPIX chip (version 2) \cite{MEDIPIX2} uses $256 \times
256$, $55 \mu$m $\times 55 \mu$m large pixels fabricated in $0.25
\mu$m technology, energy windowing via two tunable discriminator
thresholds, and a 15 bit counter. The maximum count rate per pixel
is about 1 MHz. Figure \ref{sardine} shows an X-ray scan of a
sardine taken in successive scanning steps with the MEDIPIX1 chip
($64 \times 64$ pixels, $170 \mu $m $\times 170 \mu $m) using a Si
sensor. GaAs sensors have also been successfully operated.

\begin{figure}[htb]
 \centerline{\epsfig{file=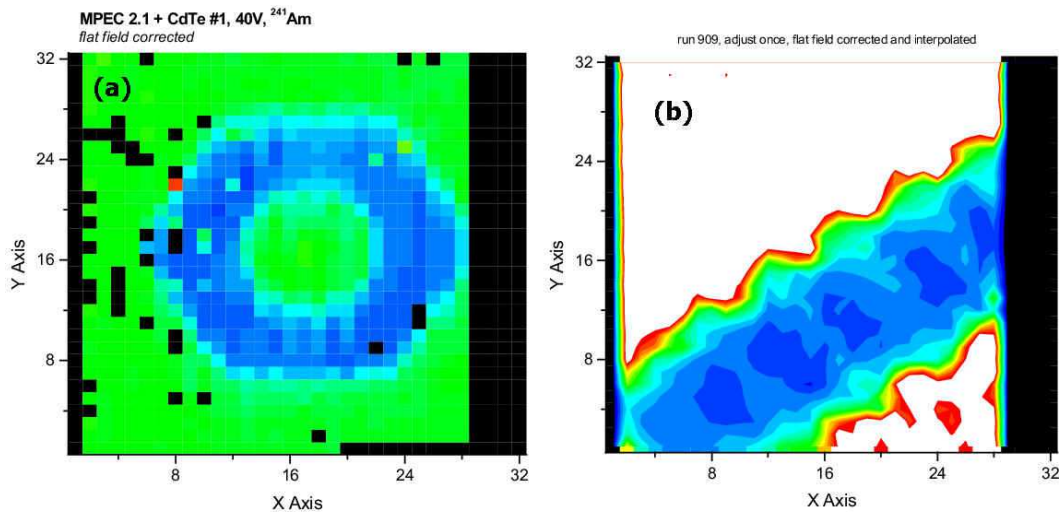,width=140mm}}
  \caption{Xray ($^{241}$Am) images, obtained with the MPEC chip ($32
\times 32$ pixels, $200 \mu $m $\times 200 \mu $m) and a CdTe
sensor, of a nut (a) and a screw in (b). The screw profile
indicates the good contrast quality obtainable by digital X-ray
imaging employing single photon counting.}
  \label{MPEC-images}
\end{figure}

Imaging sensor systems \cite{MPEC-highZ,MPEC-CdTe} using the MPEC
chip have been made also with high Z semiconductors for more
efficient X-ray absorption. A technical issue here is the bumping
of individual die sensors of Cd(Zn)Te or GaAs. This has been
solved at ISAS, Tokyo in collaboration with Mitsubishi Industries
and Bonn University by employing double Au-stud bumps with
In-filling material on order to account for thickness
inhomogeneities in the sensor surface \cite{MPEC-CdTe}. Figure
\ref{MPEC-images} shows an image of a nut and that of a screw
obtained with a CdTe sensor using the MPEC2.3 counting chip. The
MPEC chip features $32 \times 32$ pixels ($200 \mu $m $\times 200
\mu $m), double threshold operation, 18-bit counting at $\sim 1$
MHz per pixel as well as low noise values ($\sim 120$e with CdTe
sensor) and threshold dispersion ($21$e after tuning).

Both MPEC and MEDIPIX developments now address modules with a
larger detection area ($5-10 cm^2$) and high-Z sensors. While the
counting pixel approach is new and interesting for imaging
applications, it will be difficult to compete on a short time
scale with already commercially available integrating systems such
as flat panel imagers \cite{flat-panel,a-Se}. The possible market
value of photon counting pixel systems is not yet explored.

\subsection*{Crystallography}
\begin{figure}[htb]
\centerline{\epsfig{file=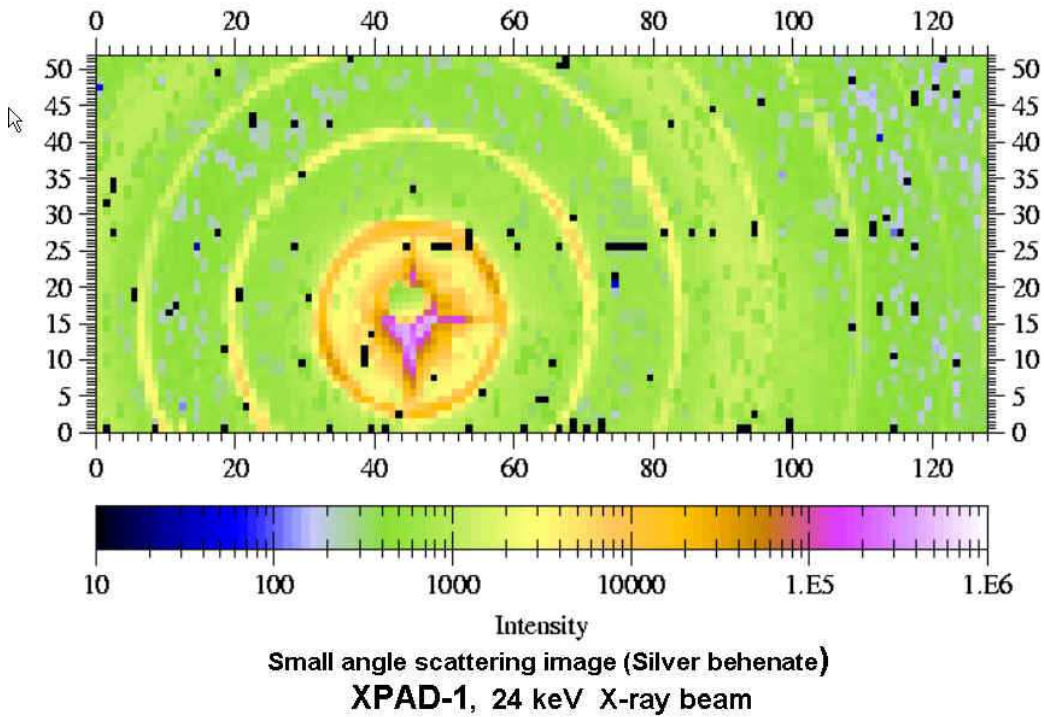,width=140mm}}
  \caption{Bragg diffraction pattern obtained with the XPAD
  hybrid counting pixel detector \cite{XPAD}.}
  \label{bragg}
\end{figure}

Counting hybrid pixel detectors are also starting to become used
in protein crystallography with synchrotron radiation. Here the
challenge is to image Bragg patterns produced by scattered Xray
photons of typically 6 keV or higher at a high rate (typ. $1-1.5$
MHz/pixel). The typical spot size of a diffraction maximum is
$100-200 \mu$m, calling for pixel sizes in the order of $100-300
\mu$m. The high linearity of the hit counting method and the
absence of so-called "blooming effects", i.e. the response of
non-hit pixels in the close neighbourhood of a Bragg maximum,
makes these detectors very appealing for protein crystallography
experiments. Developments are underway for ESRF (Grenoble, France)
and SLS (Swiss Light Source at the Paul-Scherrer Institute,
Switzerland) beam lines. The XPAD detector \cite{XPAD} ($330 \mu$m
$\times 330 \mu$m pixels) planned for ESRF beam lines has modules
with 10 chips bonded to a $4 \times 1.6$ cm$^2$ silicon substrate.
The PILATUS detector \cite{PILATUS} at the SLS ($217 \mu$m $\times
217 \mu$m pixels) is made of sixty 16 chip modules each covering
$8 \times 3.5$ cm$^2$, i.e. a total area of $40 \times 40$cm$^2$.
The maximum count rate of both detectors is $1-1.5$ MHz/pixel. A
delicate threshold tuning question remains to obtain a homogenous
response function of the detectors also at the pixel boundaries
where a loss of charge on a pixel occurs due to charge sharing and
may lead to hit losses. Figure \ref{bragg} shows a diffraction
image of Ag-Behenate obtained with the XPAD detector
\cite{delpierre}.

In conclusion, the hybrid pixel technique has been established
also for imaging using silicon and high-Z material such as GaAs or
CdTe. The method of photon counting in every pixel offers at least
in principle the possibility of an infinite dynamic range and
linearity, giving hope for a new quality of experiments with
synchrotron light detectors. For radiology the use of double
(multiple) thresholds is interesting to enhance the contrast
performance. It will, however, be very demanding to go much beyond
a count rate of about 1 MHz per pixel and to produce large area
detectors.

\section*{Monolithic Pixel Detectors}
Monolithic pixel detectors, in which amplifying and logic
circuitry as well as the radiation detecting sensor are made from
on piece of silicon, has been the dream of semiconductor detector
physicists. The first monolithic pixel detector, successfully
operated in a particle beam was made by Snoeys and collaborators
\cite{parker} as early as in 1992. They used a p-type bulk with
n-diffusion layer with ohmic contacts as pixellated collection
electrodes on the top allowing only PMOS transistors in the active
detector area. No large scale detector evolved from this approach.

A new era calling for monolithic pixel or CCD vertex detectors is
that of future colliders most notably linear $e^+e^-$ colliders,
where very low ($\ll 1\% X_0$) material per layer, small pixel
sizes ($\sim 20 \mu$m$\times 20 \mu$m) and very high rate
capability (80 hits/mm$^2$/ms) is required. This is due to the
very intense beam strahlung of a narrowly focussed electron beams
close to the interaction region which produces electron positron
pairs in vast numbers. High readout speeds of $50$ MHz with $40
\mu$s frame time are required. These challenges are addressed at
the TESLA collider \cite{TESLA} by CMOS CCDs \cite{CCD}, CMOS
Active Pixel Sensors \cite{MAPS} and DEPFET pixel detectors
\cite{DEPFET-TESLA,PFischer}.

\subsection*{CMOS Active Pixels}

\begin{figure}[htb]
\centerline{\epsfig{file=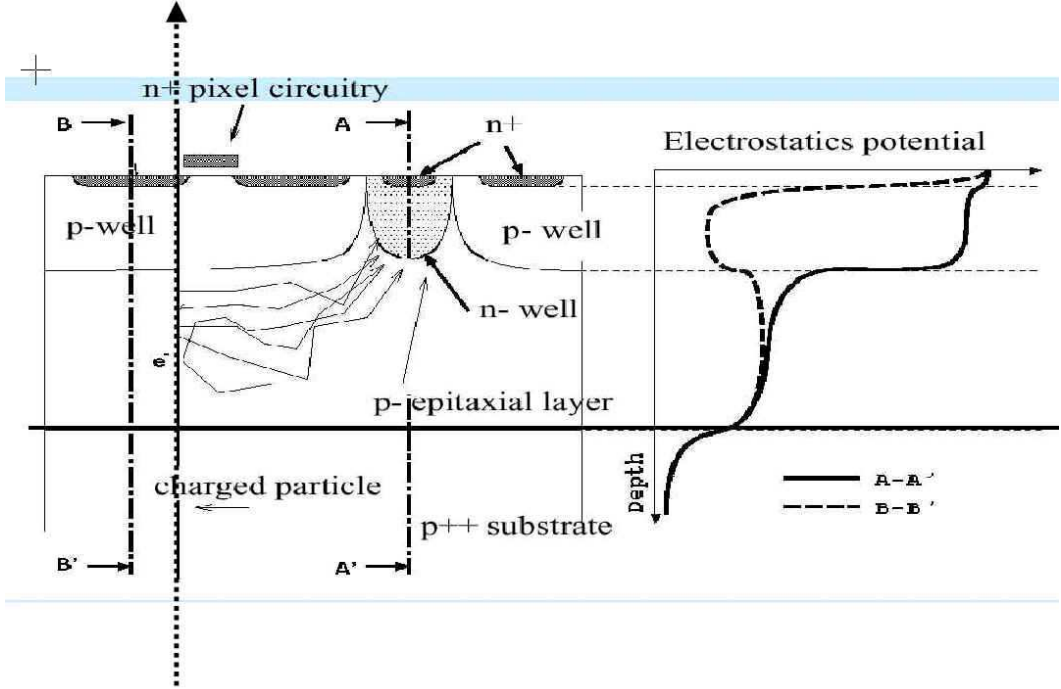,width=140mm}}
  \caption{Cross section and potential profile of the Mimosa Active
  Pixel Sensor (MAPS). Charge created in the epitaxial layer reaches the
  n-well collecting diode by diffusion.}
  \label{MAPS_structure}
\end{figure}

\begin{figure}[htb]
\centerline{\epsfig{file=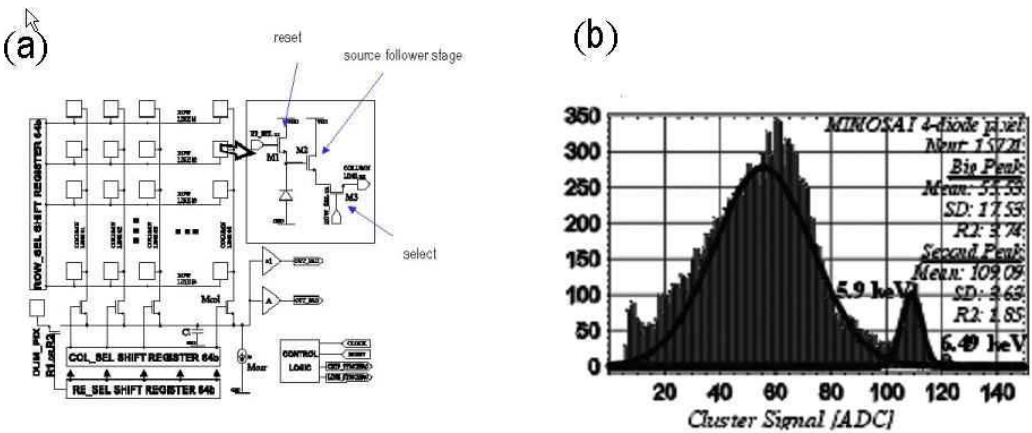,width=145mm}}
  \caption{(a) Matrix operation of MAPS-monolithic pixels. (b) Response to
  a $^{55}$Fe $5.9$ keV Xray source. An interpretation is given in the text.}
  \label{MAPS_RO}
\end{figure}

CMOS active pixel detectors as they are used in CMOS cameras are
suited also for the detection of minimum ionizing particles.
However their fill factor is less than $100\%$ and the total chip
area is small. Mimosa\footnote{minimum ionizing MOS array} Active
Pixel Sensors (MAPS) \cite{MAPS} have been developed to overcome
these deficiencies. As shown in fig. \ref{MAPS_structure} the
epitaxial layer of a standard CMOS ($0.6 \mu$m or $0.35 \mu$m)
process is used for the generation of electron-hole pairs. They
are trapped between potential barriers on both sides of the
epi-layer and reach, by thermal diffusion, an n-well/p-epi
collection diode, rendering small pixel sizes a necessity and not
a demand. The sensor is depleted only directly under this n-well
diode. Only here full charge collection is obtained. Matrix
operation is done using a standard 3-transistor readout commonly
employed by CMOS matrix devices (see fig. \ref{MAPS_RO}(a)). A
line of the matrix is selected by transistor M3, the signal is
readout via a source follower stage (M2), and reset by transistor
M1. The charge obtained for a signal from a high energy particle
is in the order of only $1000$ e or less, the collection time is
$\sim$100ns, but low noise values (15--30e) and small pixel sizes
($20 \mu$m $\times 20 \mu$m) can be achieved. This is demonstrated
in fig. \ref{MAPS_RO}(b) which shows the response to a $^{55}$Fe
$5.9$keV Xray signal. The small peak corresponds to hits entering
in the small depleted region under the n-well. The large peak is
due to Xrays absorbed in the undepleted area of a pixel.

Summarizing PRO's and CON's of CMOS monolithic pixel detectors for
particle physics I would conclude that from the above mentioned
`dream' of fully monolithic CMOS charge sensing, amplification,
readout and logic, one is still a bit away. While the low cost
(potentially $\sim 1$\$ for $4096$ pixels) for `off-the-shelf'
CMOS sensors is indeed a very attractive feature, a $100 \% $ area
fill factor still requires a special development and R$\&$D
programme (like MIMOSA APS). In the active area, due to the n-well
collection diode, no CMOS (only NMOS) circuitry is possible. The
voltage signals are very small ($\sim$mV), of the same order as
transistor threshold dispersions which at least requires some
dedicated design effort. At present \cite{Woitek} also a severe
radiation tolerance issue exists. While the MAPS sensors withstand
non-ionizing radiation (neutrons) up to about $10^{12} - 10^{13}$
n/cm$^2$, ionizing radiation, even soft X-rays, impose a serious
problem for present designs as the devices stop functioning
already after $\sim 200$krad. This is under investigation. It
seems that the radiation sensitivity is not inherently due to the
MAPS sensor structure itself. Last, but not least, the necessity
of a sufficiently thick epitaxial layer as charge collection
layer, renders only a few processing technologies suited for
sensors. With the rapid change of commercial process technologies
this also is an issue of concern. The CMOS camera market may save
these developments.

\subsection*{DEPFET Pixels}
\begin{figure}[htb]
\centerline{\epsfig{file=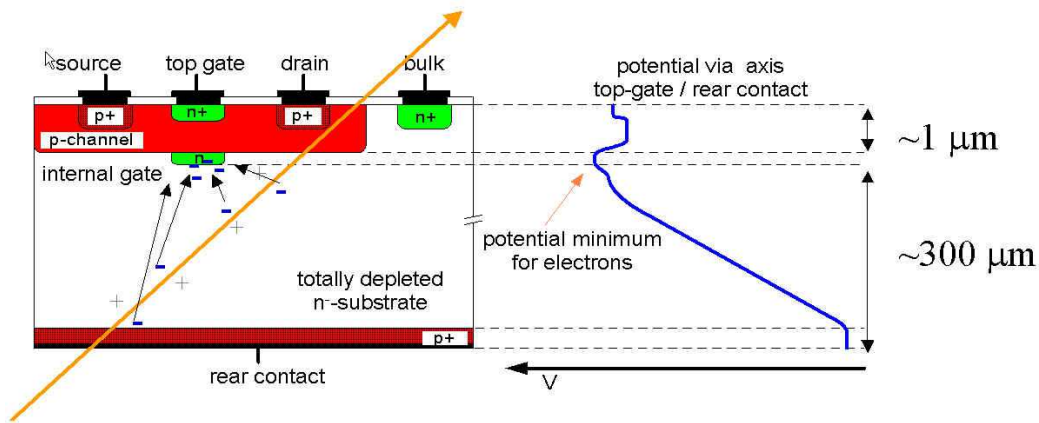, width=140mm}}
  \caption{Principle of operation of a DEPFET pixel structure.
  Cross sectional view (left) of a half pixel with symmetry axis at the
  left edge, and potential profile (right).
  }
  \label{DEPFET_principle}
\end{figure}

The DEPFET (Depletion Field Effect Transistor) pixel principle
\cite{Kemmer-Lutz} has been and is being developed for three very
different application areas: particle physics vertex detection
\cite{DEPFET-TESLA}, X-ray astronomy \cite{DEPFET-XRAY,PLechner}
and for biomedical autoradiography \cite{DEPFET-AUTORADIOGRAPHY}.
Figure \ref{DEPFET_principle} shows the principle of operation of
this device. On a sidewards depleted bulk (Si) the potential for
electrons is formed such that the minimum is located about $1
\mu$m below the top surface.
By additionally shaping the potentials of bulk, source and drain
of an implanted JFET or MOS transistor on the pixel's surface,
aided by a deep-n implantation under the gate region, a local
potential minimum for electrons is formed under the transistor
channel which acts as an ´`internal gate`. The holes created by
the impinging radiation drift towards the rear contact, the
electrons are collected and stored in the potential bucket of the
internal gate, thereby changing its potential resulting in a
modulation of the channel current of the implanted transistor. The
collected electrons are removed from the internal gate by a clear
pulse applied to a dedicated CLEAR contact outside the transistor
reagion or by other clear mechanisms e.g. through the external
gate by punch-through to the internal gate.

\begin{figure}[htb]
\centerline{\epsfig{file=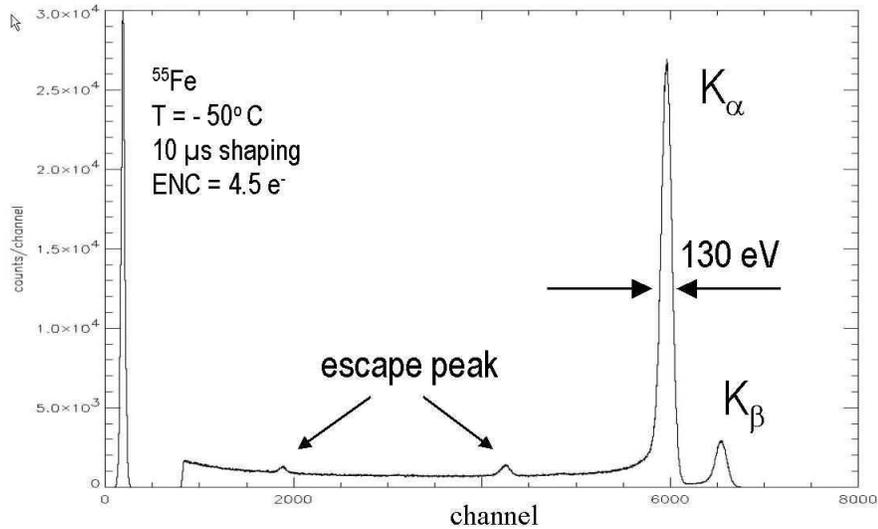,width=120mm}}
  \caption{Response of a DEPFET pixel detector to an $^{55}$Fe
  $6$keV X-ray source obtained using a single pixel device
  operated at $-50^{o}$C with a shaping time of $10 \mu$s.
  }
  \label{DEPFET_eres}
\end{figure}

The very low input capacitance and the in situ amplification (i.e.
charge to current conversion) of the device makes DEPFET pixel
detectors very attractive for low noise operation. The latter is
very important for low energy X-ray astronomy and for
autoradiography applications. For particle physics, where the
signal charge is large in comparison, this feature can be used to
design very thin detectors \cite{DEPFET-TESLA,Laci} ($\sim 30
\mu$m) with very low power consumption when operated as a row-wise
selected matrix. Depending on the application, i.e. for very low
noise operation in spectroscopy or very fast readout in particle
physics the device is operated in source follower readout mode or
drain current readout mode (see paper given at this conference
\cite{PFischer}). Figure \ref{DEPFET_eres} shows the response of a
single DEPFET pixel operated in source follower readout mode to an
$^{55}$Fe source. The measured energy resolution is
$$
\Delta {\mathrm E} = 130 \, {\mathrm eV} \quad {\mathrm @} \,\,
{\mathrm 6} \, {\mathrm keV}
$$
at a temperature of $-50 ^{o}$C. Very similar values were also
obtained at room temperature. The noise contribution is dominated
by Fano noise ($\sim 14 e$ at 6 keV and RT). The DEPFET structure
itself contributes with $4.5 e$, mostly 1/f, channel noise.

\begin{figure}[htb]
{\epsfig{file=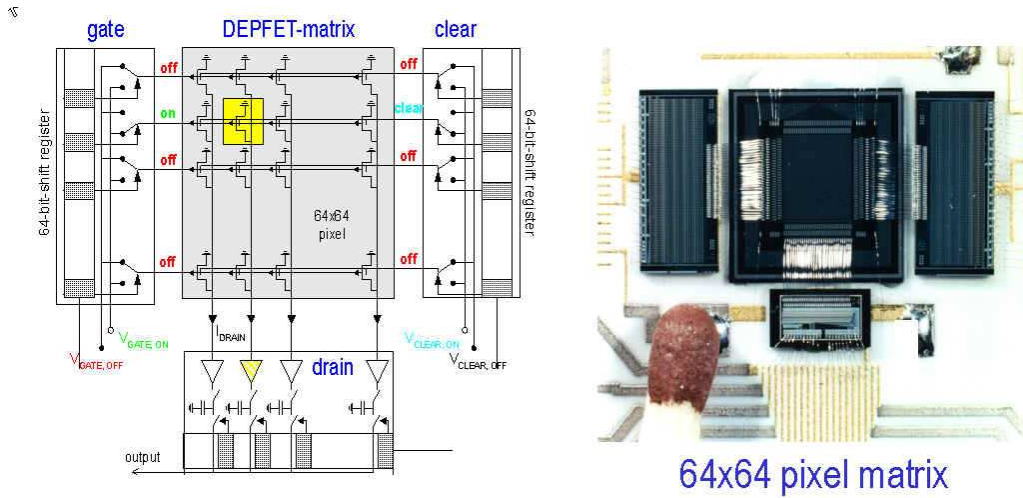,width=140mm}}
  \caption{Principle of operation (left) and photograph (right)
  of a DEPFET pixel matrix showing two steering ICs for gate and
  clear control, respectively, as well as the second current
  amplification stage at the bottom.
  }
  \label{DEPFET_matrixoperation}
\end{figure}

Figure \ref{DEPFET_matrixoperation} shows the principle of
operation of a large pixel matrix \cite{DEPFET-Ulrici}. Rows are
selected by applying a voltage to the external gate of a row.
Drains are connected column-wise. The drain current of each pixel
in a selected row is detected and amplified in a dedicated
amplification circuit. Pedestals are taken the same way several
cycles before and subtracted. Finally, clear pulses are applied to
the clear contacts to empty the internal gates.

In imaging operation spatial resolutions of
$$
27 \, {\mathrm lp / mm} \quad {\mathrm @} \quad 30 \% \,
\mathrm{MTF}
$$
have been measured \cite{DEPFET-AUTORADIOGRAPHY}.

\begin{figure}[htb]
{\epsfig{file=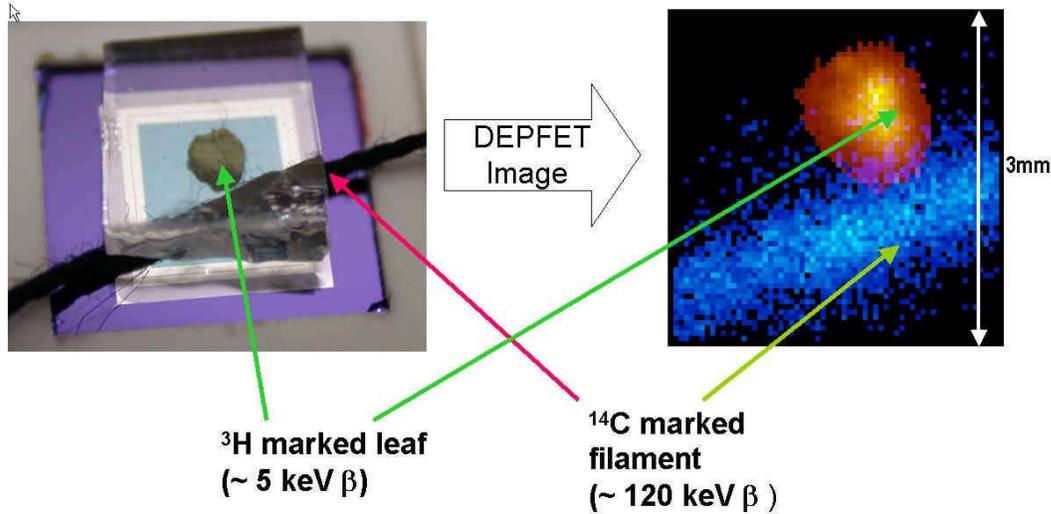,width=140mm}}
  \caption{Autoradiography image of bio-material radio-labelled
  with two different radio-markers ($^3$H and $^14$C) obtained
  with a DEPFET pixel detector.
  }
  \label{DEPFET-2label}
\end{figure}

DEPFET pixel matrices are presently being developed for low noise
high energy resolution operation in the X-ray satellite project
XEUS \cite{DEPFET-XRAY} and for high speed, high spatial
resolution in the TESLA vertex detector \cite{DEPFET-TESLA}. A
DEPFET Bioscope matrix has already been used \cite{DEPFET-Ulrici}
in autoradiographical imaging of tritium marked biological
samples. Figure \ref{DEPFET-2label} demonstrates the successful
separation of two different radio markers ($^3$H and $^{14}C$)
with high spatial resolution \cite{Ulrici}.

\section*{Summary}
Driven by the demands for high spatial resolution, high rate
particle detection in high energy physics, semiconductor pixel
detectors have also started to become exploited for imaging
applications. Hybrid pixel detectors, in which sensor and
electronic chip are separate entities, connected via bump bonding
techniques represent today's state of the art for both, particle
detection and imaging applications. Monolithic detectors, in the
form of CMOS sensors, are already used for imaging in CMOS
cameras. For the detection of high energy particles they are so
far of limited use mainly because their fill factor is smaller
than 1, i.e. less than 100\% detection coverage, and their
radiation tolerance is weak. The next generation of collider
experiments, however, will have to target monolithic pixel
detectors. Present approaches in this direction are: Monolithic
Active Pixel Sensors (MAPS) which try to overcome the limitations
of CMOS camera chips, and pn-DEPFET Pixel Detectors which are also
very attractive for (low energy) Xray astronomy and
autoradiographical imaging.

\hfill\break

\subsubsection*{Acknowledgements}
I would like to thank Woitek Dulinski for an honest discussion of
the features and technicalities of the MIMOSA Active Pixel
concept.

\end{document}